\documentclass[aps,pra,reprint,showpacs,notitlepage,superscriptaddress,twocolumn]{revtex4-2}
\usepackage{amssymb}
\usepackage{mathrsfs}
\usepackage{amsfonts}
\usepackage{graphicx}
\usepackage{amsmath}
\usepackage{dcolumn}
\usepackage{color}
\usepackage{subfigure}
\usepackage{epsfig}
\usepackage{soul}
\usepackage{float}
\usepackage[colorlinks,linkcolor=blue,citecolor=blue,hyperindex,bookmarks=false,pdfstartview=FitH]{hyperref}

\providecommand{\U}[1]{\protect\rule{.1in}{.1in}}

\newcommand{\figpanel}[2]{\hyperref[#1]{\ref*{#1}(#2)}}
\renewcommand{\eqref}[1]{\hyperref[#1]{(\ref*{#1})}}

\begin{document}

\title{Chiral Quantum Entanglement Transfer with Giant Atoms}

\author{Peng-Fei Wang}
\affiliation{College of Physics and Electronic Engineering, Hainan Normal University, Haikou 571158, People’s Republic of China}
\author{Lei Huang}
\affiliation{College of Physics and Electronic Engineering, Hainan Normal University, Haikou 571158, People’s Republic of China}
\author{Yi-Long-Yue Guo}
\affiliation{College of Physics and Electronic Engineering, Hainan Normal University, Haikou 571158, People’s Republic of China}
\author{Jing Wang}
\affiliation{College of Physics and Electronic Engineering, Hainan Normal University, Haikou 571158, People’s Republic of China}
\author{Han-Xiao Zhang}
\affiliation{College of Physics and Electronic Engineering, Hainan Normal University, Haikou 571158, People’s Republic of China}
\author{Hong Yang}
\affiliation{College of Physics and Electronic Engineering, Hainan Normal University, Haikou 571158, People’s Republic of China}
\author{Dong Yan}
\email{yand@hainnu.edu.cn}
\affiliation{College of Physics and Electronic Engineering, Hainan Normal University, Haikou 571158, People’s Republic of China}

\date{\today}

\begin{abstract}

We investigate entanglement transfer in a multi‑giant‑atom waveguide system. By tailoring chiral spontaneous emission and exploiting dark‑state dynamics, the setup enables perfect, unidirectional sequential (or selective) transfer of quantum states and their associated entanglement. The distance between two entangled atoms, i.e., the entanglement length, can be dynamically adjusted, allowing robust conversion between long‑range and short‑range entanglement during propagation. The system inherently converges to a dark state, guaranteeing high‑fidelity directional transfer. When the additional phase is modulated as a periodic piecewise function, spatially separated giant atoms exhibit stable, nearly lossless state exchange and maintain steady entanglement even under non‑Markovian conditions. This behaviour mimics conventional braided architectures without suffering from propagation delays or spatial restrictions. Our proposal offers a scalable pathway for continuous long‑distance entanglement transport and resilient state exchange in quantum networks.

\end{abstract}

\maketitle
\section{Introduction}
\label{section1}
Giant atoms are quantum emitters that interact with their environment via multiple independent coupling points~\cite{fiveyear}. This nonlocal atom–field interaction gives rise to self-interference effects that profoundly modify atomic decay and coherence~\cite{fiveyear,gustafsson2014propagating,LambAFK}, enabling phenomena such as decoherence-free (DF) interactions~\cite{braidedkannan2020waveguide,NoriGA,FCdeco,complexDFI,PhysRevA.105.023712,PhysRevA.107.013710,PhysRevResearch.6.043222,leonforte2024quantum}, chiral spontaneous emission~\cite{PhysRevX.13.021039,chen2023giant,du2022giant,wang2022chiral,wang2024nonlinear,crzs-k718}, non-Markovian retardation~\cite{andersson2019non,guo2017giant,du2022giantPRR,xu2024catch,PhysRevA.109.023712}, and unconventional bound states~\cite{PhysRevA.107.023716,WXchiral1,oscillating1}. Experimentally, giant atoms have been realized with superconducting qubits coupled to surface acoustic waves~\cite{gustafsson2014propagating,andersson2019non} and microwave transmission lines ~\cite{braidedkannan2020waveguide,PhysRevX.13.021039,vadiraj2021engineering}. Other implementations have been proposed in dynamic optical lattices~\cite{gonzalez2019engineering}, coupled waveguide arrays~\cite{longhi2020photonic,PhysRevA.107.023716}, Rydberg atoms~\cite{chen2023giant,chen2024giant,cc46-f919}, synthetic photonic dimensions~\cite{du2022giant,xiao2022bound}, and spin ensembles~\cite{wang2022giant}.

Quantum entanglement is central to quantum information science and underpins both fundamental physics and emerging technologies~\cite{PhysRev.47.777,RevModPhys.81.865,duarte2021quantum}. Over the past decades, entanglement has been generated and manipulated across diverse platforms~\cite{RevModPhys.84.777,RevModPhys.75.281,blinov2004observation,RevModPhys.73.565,walther2006cavity,RevModPhys.93.025005,PhysRevA.69.062320,wallraff2004strong}. Among these, waveguide quantum electrodynamics (QED)~\cite{RevModPhys.89.021001,gu2017microwave,RevModPhys.95.015002} has emerged as a powerful platform for long-distance entanglement generation, scalable quantum networks~\cite{kimble2008quantum} and information processing architectures~\cite{PhysRevLett.111.090502,paulisch2016universal}. However, in conventional atom–waveguide systems, state and entanglement transfer are often limited by dissipation and limited controllability, leading to information loss. Giant atoms offer a compelling alternative: their low dissipation and multi-point coupling geometry naturally suppress information loss and enhance tunability, making them promising for high-fidelity quantum state and entanglement transfer~\cite{PhysRevA.108.023728,PhysRevA.106.063703,cai2023nonreciprocal,PhysRevLett.130.053601}.

In giant-atom systems, quantum state transfer is intimately related to the generation of entanglement. A common approach to achieving this is the braided coupling configuration, which under Markovian conditions enables near-perfect state transfer via decoherence-free interactions. However, in multi-atom systems, state transfer is often not strictly sequential, which can lead to simultaneous entanglement generation. Moreover, under non-Markovian conditions, information loss can occur during transfers~\cite{braidedkannan2020waveguide,PhysRevA.107.013710}.

Building on the framework of Refs.~\cite{wang2022chiral,crzs-k718,wang2024nonlinear}, we show that in a system of giant atoms coupled to a one‑dimensional waveguide, perfect unidirectional sequential transfer of quantum states and their corresponding entanglement can be realized. This is achieved by engineering the atom‑waveguide coupling coefficients to satisfy periodic time‑reversal symmetry and by introducing additional phase modulations that enable chiral spontaneous emission and dark‑state evolution. Under these conditions, the system supports strictly sequential entanglement propagation: the entangled pair is created, annihilated, and recreated between neighboring atoms along the chain, enabling directional entanglement flow. By tuning the additional phase, selective transfer between non‑adjacent atoms becomes accessible, and the entanglement length, i.e., the distance between two entangled atoms, can be varied dynamically, allowing robust interconversion between long‑range and short‑range entanglement during propagation. 

Furthermore, when the additional phase modulation is engineered as a periodic piecewise function, two distant giant atoms exhibit persistent, nearly lossless state exchange that alternates back and forth, and they sustain stable entanglement even under non‑Markovian dynamics. This behaviour mimics that of conventional braided giant‑atom configurations but avoids their structural constraints and propagation‑delay limitations. Unlike conventional braided architectures, where long‑distance state exchange is hindered by structural complexity and non‑Markovian delays leading to reduced quantum-state transfer fidelity~\cite{braidedkannan2020waveguide,complexDFI,PhysRevA.107.013710}, our model requires only two giant atoms to achieve long‑range, near‑lossless quantum state transfer and generate stable entanglement under non‑Markovian conditions. The system spontaneously evolves into and remains in a dark state, further guaranteeing high‑fidelity directional transfer. Our scheme thus provides a scalable route towards continuous, long‑distance entanglement transport and robust state exchange in quantum networks.

In Sec.~\ref{section2}, we introduce the system model and equations of motion. Section~\ref{section3} establishes the two key conditions for achieving near-perfect and stable quantum state transfer, namely chiral spontaneous emission and dark-state construction, and analyses their close connection to entanglement dynamics. In Sec.~\ref{section4}, we demonstrate unidirectional sequential transfer of quantum states and entanglement, and show that by tuning the additional phase, selective transfer between nonadjacent atoms becomes accessible. Moreover, the entanglement length (the distance between two entangled atoms) can be dynamically varied, enabling robust interconversion between long‑range and short‑range entanglement during propagation. Finally, Sec.~\ref{section5} reveals the mechanism underlying sustained state exchange and steady entanglement under a periodic piecewise phase modulation, which reverses the chiral direction periodically and produces persistent, nearly lossless back‑and‑forth state exchange even under non‑Markovian dynamics. Sec.~\ref{section6} concludes the paper.

\section{Model and dynamical equations}
\label{section2}

\begin{figure}[htbp]
  \centering
    \vspace{0pt}
    \includegraphics[width=1\linewidth]{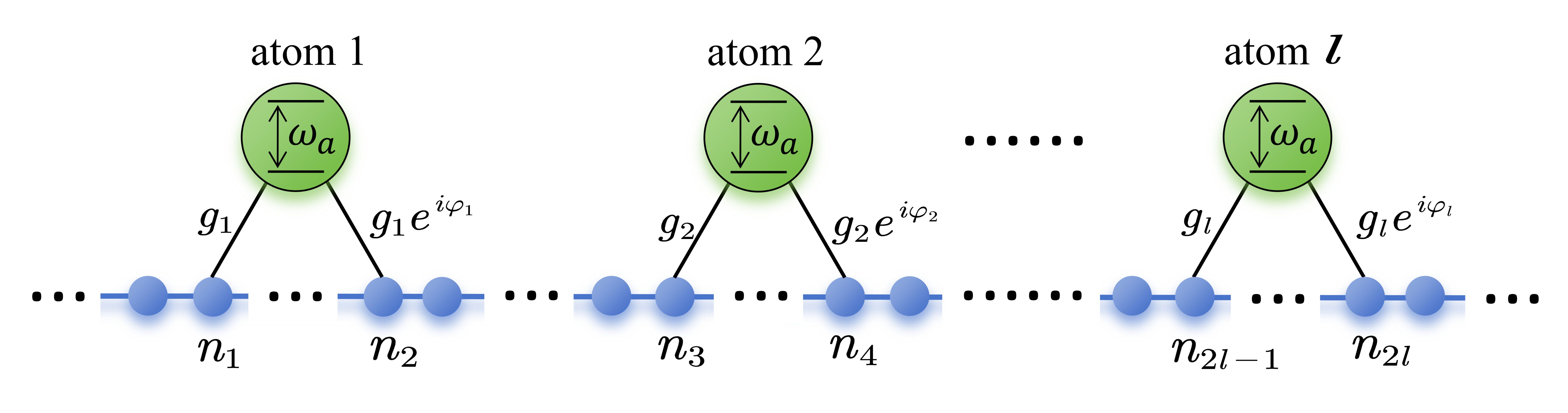}
    \caption{Schematic of a discrete multi-giant-atom structure coupled to an infinite one-dimensional lattice. Each two-level giant atom labeled $l$ has transition frequency $\omega_a$. The $l$-th atom couples to the lattice sites $n_{2l-1}$ and $n_{2l}$ with time-dependent coefficient $g_l(t)$. The atom size $d_s=n_{2l}-n_{2l-1}$ and the interatomic distance $d_w=n_{2l+1}-n_{2l}$ are uniform across the array. An additional phase difference $\varphi_l$ between the two coupling points of each atom enables control of interference effects such as chiral emission and dark-state engineering.} 
    \label{FIG1}
\end{figure}

Figure~\ref{FIG1} shows the system: an array of $N$ identical two-level giant atoms with transition frequency $\omega_a$ coupled to an infinite one-dimensional lattice. Each lattice site has resonance frequency $\omega_0$, and $\eta$ denotes the nearest-neighbor hopping rate. The $l$-th atom couples to the lattice sites $n_{2l-1}$ and $n_{2l}$ with the time-dependent coupling coefficient $g_l(t)$. A phase difference $\varphi_l$ between the two coupling points breaks the time-reversal symmetry. The total Hamiltonian is $H=H_a+H_w+H_{aw}$ ($\hbar= 1$), 

\begin{subequations}
\begin{align}
H_a &= \sum_{l=1}^N \omega_a \sigma_{l}^{+} \sigma_{l}^{-}, \\
H_w &= \sum_{j} \omega_0 a_{j}^{\dagger} a_j
- \sum_{j} \eta \left( a_j a_{j+1}^{\dagger} + \mathrm{H.c.} \right), \\
H_{aw} &= \sum_{l=1}^N g_l(t)
\left[\sigma_{l}^{-}
\left(a_{n_{2l-1}}^{\dagger}
+ e^{i \varphi_l} a_{n_{2l}}^{\dagger} \right)
+ \mathrm{H.c.} \right].
\end{align}
\label{EQ1}
\end{subequations}

\noindent 
Here, $a_j$ ($a_j^\dagger$) annihilates (creates) a photon at site $j$, and
$\sigma_l^- = |g\rangle_l {}_l\langle e|$
($\sigma_l^+ = |e\rangle_l {}_l\langle g|$) is the lowering (raising) operator of atom $l$ with the ground state $|g\rangle_l$ and the excited state $|e\rangle_l$.

In the single-excitation subspace, the system state is

\begin{equation}
|\Psi (t)\rangle =\left[\sum_j{c_j(t)a_{j}^{\dagger}}+\sum_{l=1}^N{u_l(t)\sigma _{l}^{+}} \right] |G\rangle,
\label{EQ2}
\end{equation}
where $c_j(t)$ and $u_l(t)$ denote the excitation amplitudes of the $j$-th lattice site and the $l$-th atom at time $t$, respectively, and $|G\rangle={|g \rangle}^{\otimes N}{\otimes}{|\emptyset \rangle}^{\otimes \infty}$ ($|\emptyset \rangle$ is the vacuum state of a lattice site). The Schr\"odinger equation yields

\begin{equation}
\begin{split}
\dot{u}_l(t) &= -i\omega_a u_l(t) - i g_l(t) \left[ c_{n_{2l-1}}(t) + e^{-i\varphi_l} c_{n_{2l}}(t) \right], \\
\dot{c}_j(t) &= -i \omega_0 c_j(t) + i \eta \left[c_{j-1}(t) + c_{j+1}(t)\right] \\
&\quad - i \sum_{l=1}^N g_l(t) u_l(t) \left( \delta_{j,n_{2l-1}} + e^{i\varphi_l} \delta_{j,n_{2l}} \right).
\end{split}
\label{EQ3}
\end{equation}

\section{Fundamental principles for realizing chiral entanglement}
\label{section3}
\subsection{Chiral Spontaneous Emission}
\label{section3a}

In the interaction picture, the Hamiltonian~\eqref{EQ1} becomes
\begin{equation}
\begin{split}
H_\mathrm{I} =\sum_{l=1}^N{\left[ \sum_k{g_{l}^{k}\left( t \right) e^{i\left( \omega _k-\omega _a \right) t}
a_{k}^{\dagger}\sigma _{l}^{-}}+\mathrm{H}.\mathrm{c}. \right]},
\end{split}
\label{EQ4}
\end{equation}
with $g_{l}^{k}\left( t \right) =\frac{g_l(t)}{\sqrt{M}}e^{-ikn_{2l-1}}\left[ 1+e^{-i\phi_l(k)}e^{i\varphi _l} \right]$, $\omega_k = \omega_0 - 2\eta \cos(k)$, and $\phi_l(k)=kd_s$ with $k$ the photon wave vector. The squared coupling amplitude is~\cite{roccati2024controlling}
\begin{equation}
\begin{split}
|g_{l}^{k}(t)|^2=\frac{2\left[ g_l(t)\right]^2}{M}\left[1+\cos(-\phi_l(k)+\varphi_l)\right].
\end{split}
\label{EQ5}
\end{equation}
The group velocity
\begin{equation}
v_g = 2 \eta \sin(k)
\label{EQ6}
\end{equation}
governs not only the speed of propagation but also, crucially for our purposes, its directionality.

Setting $\omega_a \equiv \omega_0$, we analyse the lattice energy band relative to the giant-atom transition frequency. From Eq.~\eqref{EQ6}, the photon modes supported under this condition are left- and right-propagating waves with wave vectors $k_{\mathrm L}=-\pi/2$ and $k_{\mathrm R}=\pi/2$, respectively. According to Eq.~\eqref{EQ5}, chiral spontaneous emission into a given direction requires coupling to the opposite mode to vanish while that to the desired mode remains finite. Specifically, emission into the right-propagating mode and the complete suppression of left-propagating emission are achieved when $-\phi _l(k_{\mathrm L})+\varphi_l = \pi+2n\pi$ and $-\phi _l(k_{\mathrm R})+\varphi_l \neq \pi+2n\pi$; conversely, emission into the left-propagating mode requires $-\phi _l(k_{\mathrm R})+\varphi_l = \pi+2n\pi$ and $-\phi _l(k_{\mathrm L})+\varphi_l \neq \pi+2n\pi$.

\begin{figure}[htbp]
  \centering
    \vspace{0pt}
    \includegraphics[width=1\linewidth]{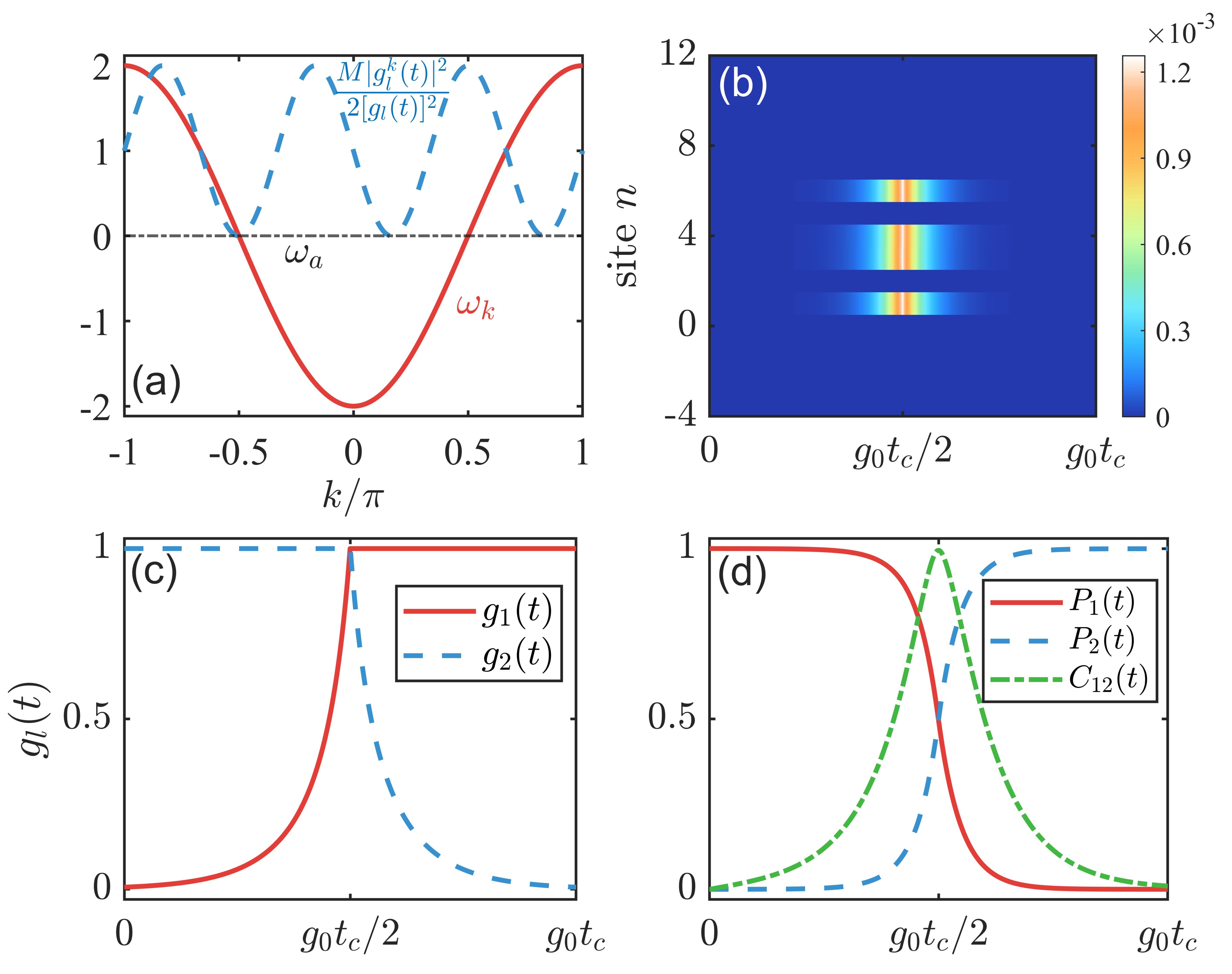}
    \caption{(a) The relationship among the lattice band frequency $\omega_k$, atomic frequency $\omega_a$, and $\frac{M|g_{l}^{k}(t)|^2}{2[g_l(t)]^2}$. (b) Time evolution of the field intensity distribution $|c_j(t)|^2$. (c) Time evolution of the coupling coefficients $g_l(t)$. (d) Time evolutions of the population $P_l(t)=\left| u_l(t) \right|^2$ and the concurrence $C_{12}(t)$. Other parameters are $N=2$, $|\Psi (0)\rangle =|\psi_1\rangle$, $\alpha = 0.044$, $g_0t_0=200$, $g_0\tau_{sw}=0.1$, $g_0 t_c=200.1$, $\eta/g_0=20$, $n_1=0$, $n_2=3$, $n_3=4$, $n_4=7$, $d_s=3$, $d_w=1$, and $\varphi_1=\varphi_2=-\pi/2$.}
    \label{FIG2}
\end{figure}

\subsection{Entangling atoms via dark-state design}
\label{section3b}
The jump operators associated with the right- and left-propagating modes are given by~\cite{wang2022chiral,roccati2024controlling,NoriGA,PhysRevA.105.023712,gough2009quantum,combes2017slh}:
\begin{subequations}
\begin{align}
J_{L}^{N}&=\sum_{l=1}^N{e^{ik_{a}(l-1)d_{sw}}\sqrt{\Gamma _{L}^{l}(t)}\sigma _{l}^{-}},
\label{EQ7a}\\
J_{R}^{N}
&=\sum_{l=1}^N e^{ik_{a}(N-l)d_{sw}}\sqrt{\Gamma_{R}^{l}(t)}\sigma _{l}^{-},
\label{EQ7b}
\end{align}
\label{EQ7}
\end{subequations}

\noindent 
where $k_a=\pi/2$, $d_{sw}=d_{s}+d_{w}$, $\Gamma_{\mathrm R}^{l}\left(t\right) =\Gamma _l\left(t\right)\left[1+\cos\left( -\phi_l(k_\mathrm R)+\varphi_l\right) \right]/2$, $\Gamma_{\mathrm L}^{l}\left(t\right) =\Gamma_l\left(t\right)\left[1+\cos \left(-\phi_l(k_\mathrm L)+\varphi _l\right)\right]/2$, and $\Gamma _l(t)=4g_l^2(t)/v_g$.

When only right-propagating photons are present in the waveguide, we have $J_L^{N}=0$. Under this condition, the propagation phase $\phi_l(k_{\mathrm L})$ and the coupling phase $\varphi_l$ must satisfy $-\phi_l(k_{\mathrm L})+\varphi_l=\pi+2n\pi$ for some integer $n$. This phase-matching condition ensures that the dark-state evolution requires the right-propagating jump operator to satisfy $J_{\mathrm R}^{N}|\Psi(t)\rangle=0$, which imposes the following constraint on the probability amplitudes:

\begin{equation} 
\sum_{l=1}^N e^{i\pi(N-l)d_{sw}/2}\sqrt{\Gamma_l(t)}\left|\sin\left[\phi_l\left(k_R\right)\right]\right|u_l(t)=0. 
\label{EQ8}
\end{equation}
Note that equation~\eqref{EQ8} simplifies to a very simply form when $d_{sw}=4n$ with $n$ an integer. 

For a two-atom system, setting $d_s=3$, $d_w=1$, and $\varphi_l=-\pi/2$, Eq.~\eqref{EQ8} reduces to 
\begin{equation} 
\sqrt{\Gamma _1(t)}u_1(t)+\sqrt{\Gamma _2(t)}u_2(t)=0.
\label{EQ9}
\end{equation}
Perfect state transfer requires boundary conditions:
\begin{equation}
\begin{split}
&\left|u _1\left(0\right)\right|=\left|u _2\left(t_c\right)\right|=1,
\\
&\left|u _1\left(t_c\right)\right|=\left| u _2\left(0\right)\right|=0.
\end{split}
\label{EQ10}
\end{equation}
To simultaneously satisfy Eqs.~\eqref{EQ9} and~\eqref{EQ10}, we set the time-reversed coupling $g_1(t)=g_2(t_0+\tau_{sw}-t)$, where $\tau_{sw}=d_{sw}/v_g$ characterizes the time delay induced by non-Markovian effect. Following Refs.~\cite{wang2022chiral,crzs-k718}, we choose
\begin{subequations}
\begin{align}
g_1(t) &=
\begin{cases}
g_0\displaystyle\frac{e^{\alpha \left( t-\frac{t_0}{2} \right)}}{2-e^{\alpha \left( t-\frac{t_0}{2} \right)}},
& \ \ \ \ \ 0\leqslant t<\frac{t_0}{2},\\[6pt]
g_0,
& \ \ \ \ \ \frac{t_0}{2} \leqslant t<t_c,
\end{cases}
\label{EQ11a}
\\[10pt]
g_2(t) &=
\begin{cases}
g_0,
& 0 \leqslant t< \frac{t_0}{2}+\tau_{sw},\\[6pt]
g_0\frac{e^{-\alpha \left[ t-\left( \frac{t_0}{2}+\tau _{sw} \right) \right]}}{2-e^{-\alpha \left[ t-\left( \frac{t_0}{2}+\tau _{sw} \right) \right]}},
& \frac{t_0}{2}+\tau_{sw}\leqslant t<t_c.
\end{cases}
\label{EQ11b}
\end{align}
\label{EQ11}
\end{subequations}

\noindent 
Here $g_0$ is the maximum coupling, and $\alpha$ controls the coupling variation. 

\begin{figure}[htbp]
  \centering
    \vspace{0pt}
    \includegraphics[width=1\linewidth]{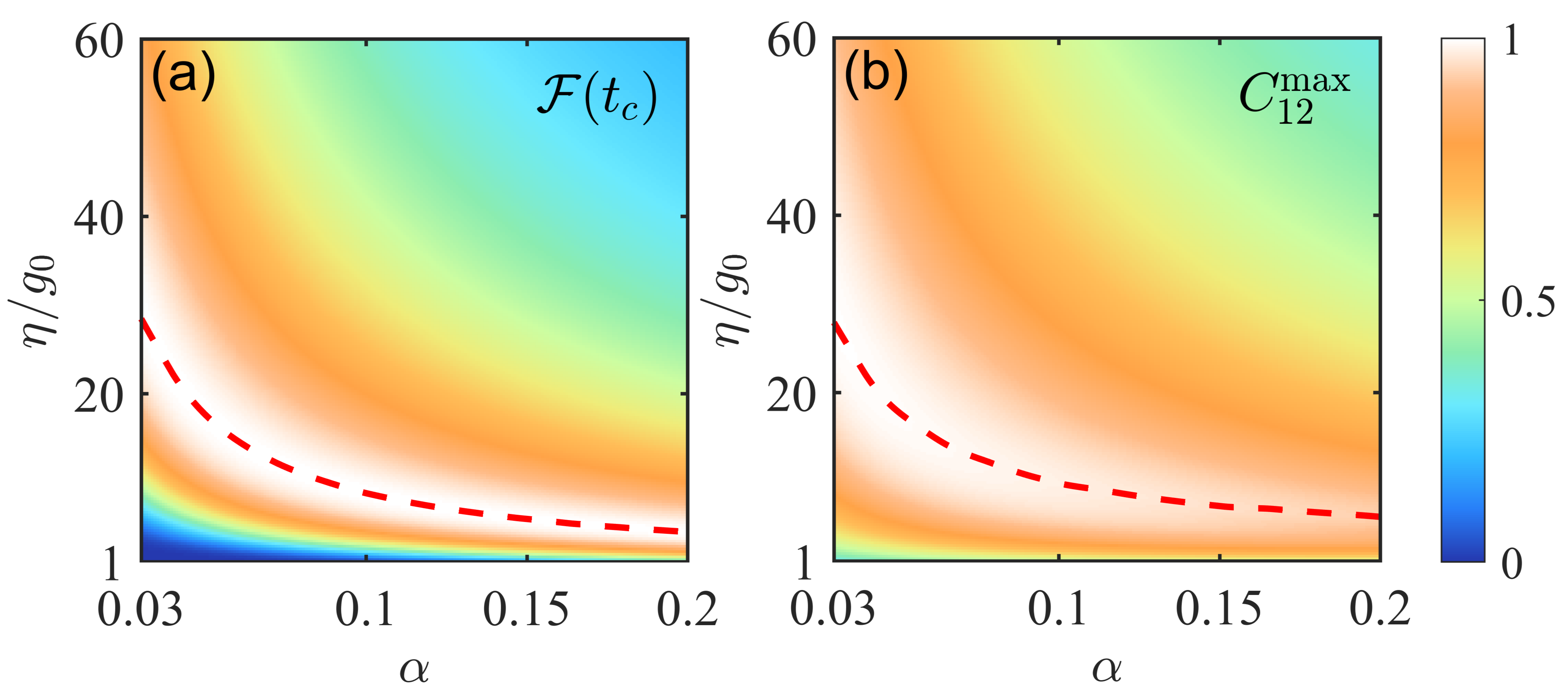}
    \caption{(a) Fidelity $\mathcal{F}(t_c)$ and maximum concurrence  $C_{12}^{\max}$ (b) as functions of the coupling coefficient $\eta$ and the parameter $\alpha$. Dashed lines indicates $\mathcal{F}(t_c)=1$ and $C_{12}^{\max}=1$, respectively. Other parameters are as in Fig.~\ref{FIG2}.}
    \label{FIG3}
\end{figure}
In the single-excitation subspace, the concurrence is~\cite{PhysRevA.106.063703,PhysRevA.108.023728,PhysRevLett.80.2245}
\begin{equation}
\begin{split}
C_{ij}(t)=2\left| u_i(t)u_{j}^{*}(t) \right|.
\end{split}
\label{EQ12}
\end{equation}

Figures~\figpanel{FIG2}{b} and~\figpanel{FIG2}{d} demonstrate complete quantum state transfer from $|\psi_1\rangle$ with $P_1(0)$=1 to $|\psi_2\rangle$ with $P_2(t_c)=1$ after a critical time $t_c=t_0+\tau_{sw}$, as well as the generation of the maximally entangled state $(|\psi_1\rangle- |\psi_2\rangle)/\sqrt{2}$ at the intermediate time $t_c/2$, using $g_1(t)$ and $g_2(t)$ shown in Fig.~\figpanel{FIG2}{c}. In general, however, the maximally entangled state does not appear exactly at $t_c/2$, but rather slightly deviates from this point.

Figure~\ref{FIG3} shows that the final fidelity $\mathcal{F}(t_c)=\left| \left\langle \Psi(t_c) \middle| \psi_2 \right\rangle \right|$ and the maximum concurrence $C_{12}^{\mathrm{max}}$ exhibit nearly identical dependence on $\eta$ and $\alpha$, indicating that the fidelity of state transfer directly determines the maximally achievable entanglement. To ensure transmission efficiency, both the dark-state condition and the adiabatic condition must be satisfied. The latter requires the mixing angle $\theta=\tan^{-1}(g_1(t)/g_2(t))$ to vary sufficiently slowly (equivalently, the integral $S=\int \sqrt{g_1^2(t)+g_2^2(t)} dt$ to be large enough), which demands a small $\alpha$. For large $\alpha$, the adiabatic condition breaks down; reducing $\eta$ then increases the relaxation rate $\Gamma_l$ into the waveguide, thereby matching the faster switching between $g_1(t)$ and $g_2(t)$. Conversely, for sufficiently small $\alpha$, the variation of $\theta$ is slow enough ($S$ nearly saturated) that the adiabatic condition is naturally met, rendering the transmission efficiency largely independent of $\eta$.

\section{Perfect Selective Entanglement Transfer}
\label{section4}

\subsection{Entanglement Transfer For An Initial Separable States}
\textbf{\label{section4a}}

To achieve selective entanglement generation and perfect transfer, we redefine the coupling coefficient $g_l(t)$. Generalizing from the time-reversed coupling Eq.~\eqref{EQ11} that generates perfect entangled states, we introduce a new time-reversal-symmetry form of $g_l(t)$ as follows~\cite{wang2022chiral,crzs-k718}
\begin{widetext}
\begin{subequations}
\begin{align}
g_a(t)=g_a\!\left(t+2nt_c\right)
&=
\begin{cases}
g_0\dfrac{e^{\alpha\left(t-\frac{t_0}{2}\right)}}{2-e^{\alpha\left(t-\frac{t_0}{2}\right)}},
& 0\leqslant t<\dfrac{t_0}{2}, \\[10pt]
g_0,
& \dfrac{t_0}{2}\leqslant t<{t_c}, \\[6pt]
g_0,
& {t_c}\leqslant t<\dfrac{3t_0}{2}+2\tau_{sw}, \\[6pt]
g_0\dfrac{e^{-\alpha\left[t-\left(\frac{3t_0}{2}+2\tau_{sw}\right)\right]}}
{2-e^{-\alpha\left[t-\left(\frac{3t_0}{2}+2\tau_{sw}\right)\right]}},
& \dfrac{3t_0}{2}+2\tau_{sw}\leqslant t<2t_c,
\end{cases}
\label{EQ14a}
\\[12pt]
g_b(t)=g_b\!\left(t+2nt_c\right)
&=
\begin{cases}
g_0,
& 0\leqslant t<\dfrac{t_0}{2}+\tau_{sw}, \\[6pt]
g_0\dfrac{e^{-\alpha\left[t-\left(\frac{t_0}{2}+\tau_{sw}\right)\right]}}
{2-e^{-\alpha\left[t-\left(\frac{t_0}{2}+\tau_{sw}\right)\right]}},
& \dfrac{t_0}{2}+\tau_{sw}\leqslant t<t_c, \\[10pt]
g_0\dfrac{e^{\alpha\left[t-\left(\frac{3t_0}{2}+\tau_{sw}\right)\right]}}
{2-e^{\alpha\left[t-\left(\frac{3t_0}{2}+\tau_{sw}\right)\right]}},
& t_c\leqslant t<\dfrac{3t_0}{2}+\tau_{sw}, \\[10pt]
g_0,
& \dfrac{3t_0}{2}+\tau_{sw}\leqslant t<2t_c,
\end{cases}
\label{EQ14b}
\end{align}
\label{EQ14}
\end{subequations}
\end{widetext}
where $n$ is an integer. Equation~\eqref{EQ14} shows that for $0<t<t_c$, the temporal evolution of $g_a(t)$ and $g_b(t)$ are identical to those of $g_1(t)$ and $g_2(t)$ in Eq.~\eqref{EQ11}, respectively. In the interval $t_c<t<2t_c$, their roles are swapped relative to the first interval. This pattern repeats with a period $2t_c$, thereby producing a periodic modulation of the couplings.

\begin{figure}[htbp]
  \centering
    \vspace{0pt}
    \includegraphics[width=1\linewidth]{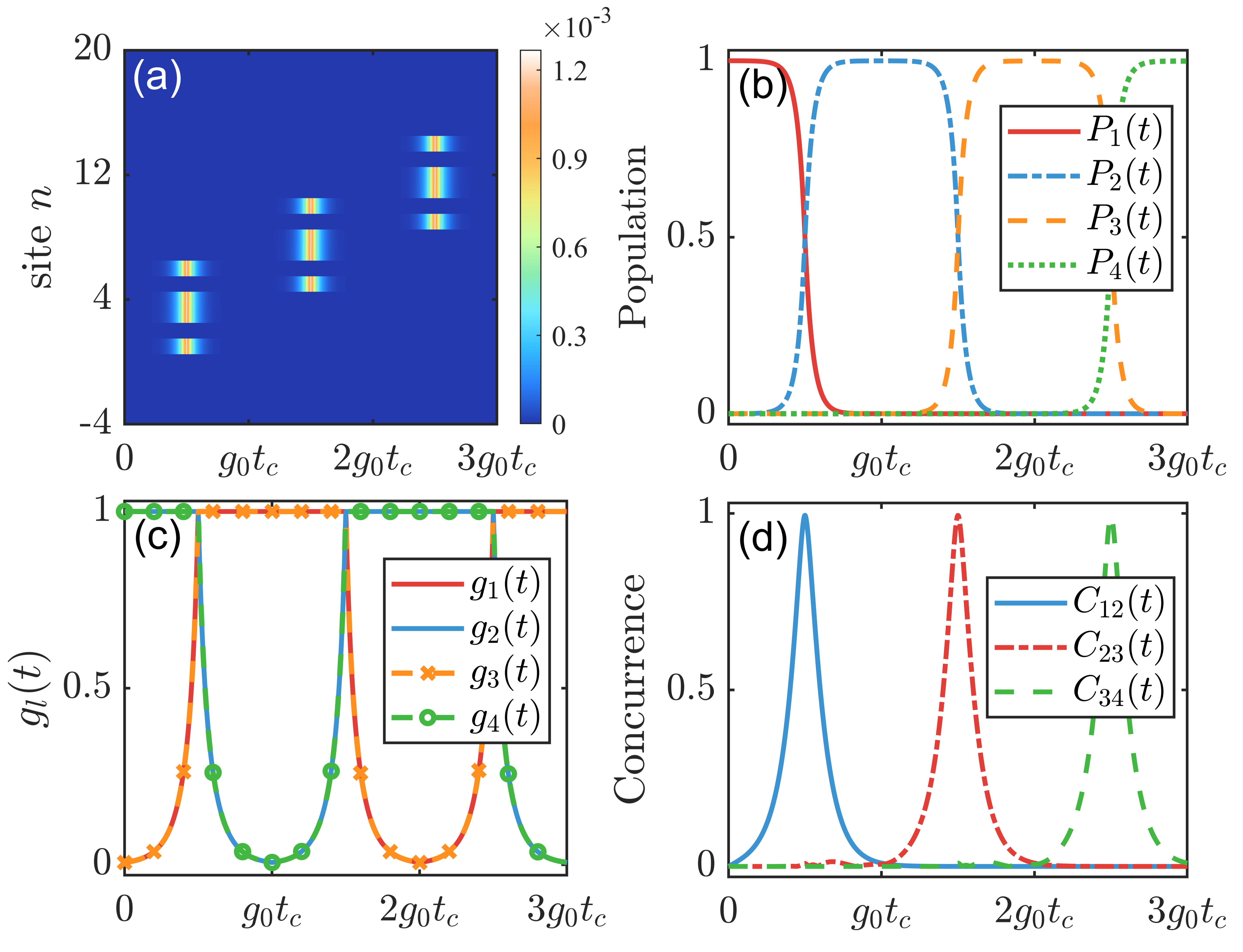}
    \caption{(a) Time evolution of the field intensity distribution $|c_j(t)|^2$. (b) Time evolution of the population $P_l(t)$ associated with state $|\psi_l\rangle$. (c) Time evolution of the coupling coefficients $g_l(t)$, with $g_{1,3}(t)=g_a(t)$ and $g_{2,4}(t)=g_b(t)$. (d) Time evolution of the concurrence between neighboring atoms. Other parameters are $N=4$, $|\Psi (0)\rangle =|\psi_1\rangle$, $g_0t_0=200$, $g_0\tau_{sw}=0.1$, $g_0 t_c=200.1$, $\eta/g_0=20$, $n_{2l-1}=4l-4$, $n_{2l}=4l-1$, $d_s=3$, $d_l=1$, and $\varphi_l=-\pi/2$.}
    \label{FIG4}
\end{figure}

Figure.~\ref{FIG4} demonstrates perfect entanglement generation and its sequential transfer via right-chiral spontaneous emission. In this case, the coupling coefficients are set as $g_{1,3}(t)=g_a(t)$ and $g_{2,4}(t)= g_b(t)$ (see Fig.~\figpanel{FIG4}{c}). Starting from $\left|\psi _1\right> $, the dynamics follow Sec.~\ref{section3} and Eq.~\eqref{EQ9}. For $0 < g_0t < g_0t_c$, atoms 1 and 2 form a dark state, driving $\left|\psi_1\right> \rightarrow \left|\psi_2\right>$. Owing to chiral directionality, during $g_0t_c<g_0t<2g_0t_c$, atom 2 preferentially forms a dark state with atom 3, yielding $\left|\psi_2\right> \rightarrow \left|\psi_3\right>$. Over $0 < g_0t < 3g_0t_c$, the evolution proceeds sequentially as $\left|\psi_1\right> \rightarrow \left|\psi_2\right> \rightarrow \left|\psi_3\right> \rightarrow \left|\psi_4\right>$ (see Figs.~\hyperref[FIG4]{4(a)} and \hyperref[FIG4]{(b)}). Each transfer cycle creates and then annihilates entanglement between the involved pair. (see Fig.~\figpanel{FIG4}{d}). Entanglement between neighboring atoms propagates sequentially as $C_{12} \rightarrow C_{23} \rightarrow C_{34}$ closely tracking the state transfer. Note that $C_{23}$ does not vanish exactly during $0 < g_0t < g_0t_c$ because that $\left|\psi _1\right> \rightarrow \left|\psi _2\right>$ is incomplete, so a small $\left|\psi _1\right> \rightarrow \left|\psi _3\right>$ transition then yields a small $C_{23}$ via Eq.~\eqref{EQ12}. The same mechanism gives small nonzero $C_{34}$ during $g_0t_c < g_0t < 2g_0t_c$.

We then set $g_{1,2,5,6}(t) = g_a(t)$, $g_{3,4,7}(t) = g_b(t)$ and phases $\varphi_{1,3,5,7} = -\varphi_{2,4,6}= -\pi/2$ for a seven-atom system. From Eq.~\eqref{EQ5}, atoms 1,3,5,7 emit only right-propagating photons, while atoms 2,4,6 emit only left-propagating photons. Under these conditions, the left- and right-jump operators become
\begin{subequations}
\begin{align}
J_{\mathrm{R}}^{7}=\sum_{l=1,3,5,7}{\sqrt{\Gamma _{R}^{\left( l \right)}\left( t \right)}\sigma _{-}^{\left( l \right)}},\\
J_{\mathtt{L}}^{7}=\sum_{l=2,4,6}{\sqrt{\Gamma _{L}^{\left( l \right)}\left( t \right)}\sigma _{-}^{\left( l \right)}}.
\end{align}
\end{subequations}

\begin{figure}[htbp]
  \centering
    \vspace{0pt}
    \includegraphics[width=1\linewidth]{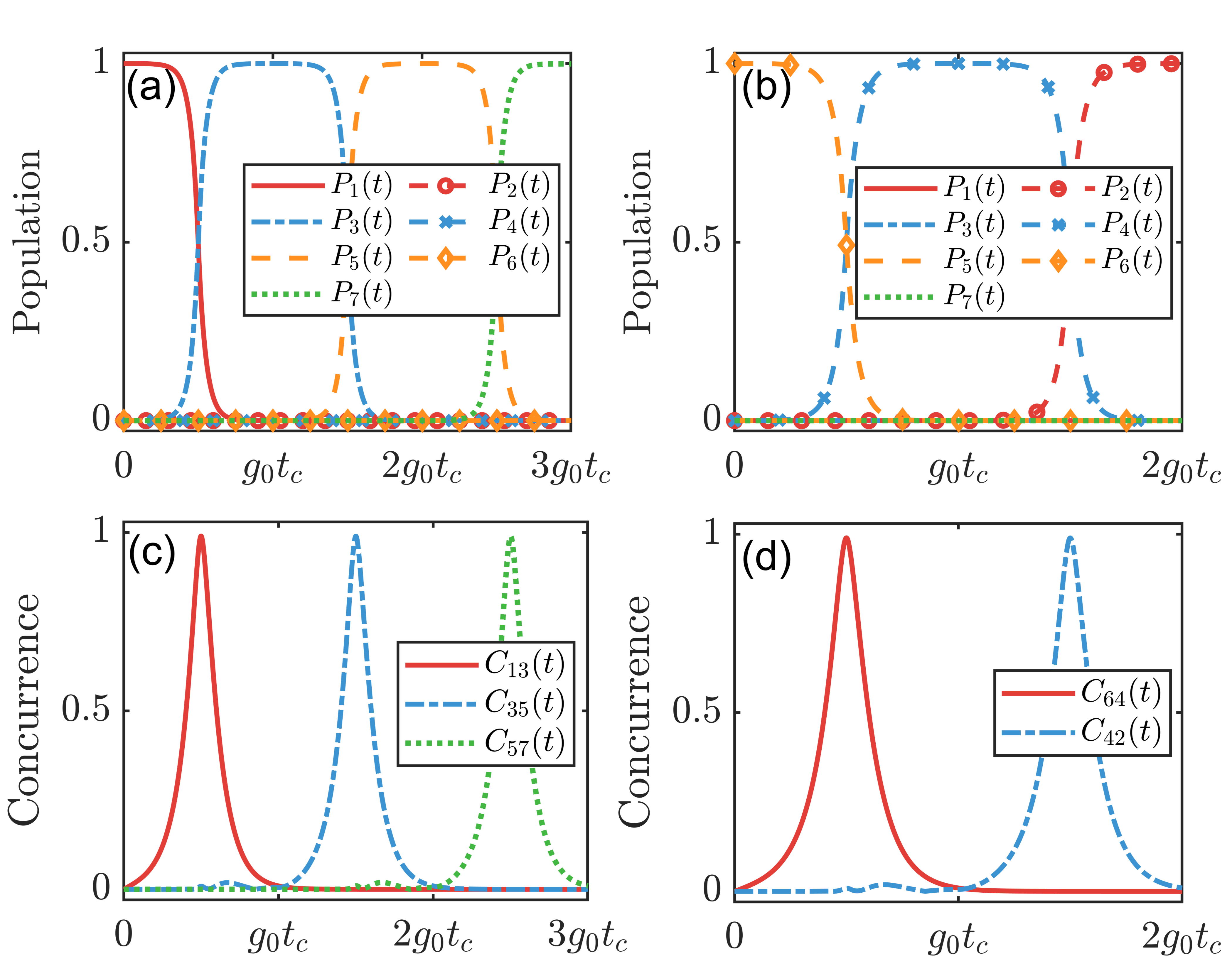}
    \caption{(a) and (b) Time evolution of the populations $P_l(t)$ for different initial states. (c) and (d) Time evolution of the concurrence between non-neighboring atoms for different initial states.  Initial state $|\Psi (0)\rangle =|\psi_1\rangle$ in (a) and (c), and $|\Psi (0)\rangle =|\psi_6\rangle$ in (b) and (d). Other parameters are $N=7$, $g_0t_0=200$, $g_0\tau_{sw}=0.1$, $g_0 t_c=200.1$, $\eta/g_0=20$, $n_{2l-1}=4l-4$, $n_{2l}=4l-1$, $d_s=3$, $d_l=1$, $\varphi_{1,3,5,7}= -\varphi_{2,4,6}=-\pi/2$,  $g_{1,2,5,6}(t) = g_a(t)$, and $g_{3,4,7}(t) = g_b(t)$.}
    \label{FIG5}
\end{figure}

These operators act on the states $|\psi_1\rangle$, $|\psi_3\rangle$, $|\psi_5\rangle$, $|\psi_7\rangle$ and $|\psi_2\rangle$, $|\psi_4\rangle$, $|\psi_6\rangle$, respectively. For the initial state $|\psi_1\rangle$ (see Figs.~\hyperref[FIG5]{5(a)} and \hyperref[FIG5]{(c)}), only right-propagating photons exist, so transitions to $|\psi _2\rangle$, $|\psi _4\rangle$, $|\psi_6\rangle$ are forbidden. During $0 < g_0t < g_0t_c$, the dark-state condition first holds for atoms 1 and 3, driving $|\psi_1\rangle \rightarrow |\psi_3\rangle$ and generating concurrence $C_{13}$. The system then sequentially transfers $|\psi_3\rangle \rightarrow |\psi_5\rangle \rightarrow |\psi_7\rangle$, with entanglement propagating as $C_{13} \rightarrow C_{35} \rightarrow C_{57}$. Conversely, for $|\psi_6 \rangle$ as the initial state (see Figs.~\hyperref[FIG5]{5(b)} and \hyperref[FIG5]{(d)}), system evolution is confined to the subspace $\{|\psi _2\rangle$, $|\psi _4\rangle$, $|\psi _6\rangle\}$, yielding perfect left-propagation and an analogous entanglement chain $C_{64} \rightarrow C_{42}$.

% Without dissipation and with suitable parameters, near‑perfect state and entanglement transfer is achieved (see Fig.~\ref{FIG4}). Including dissipation (see Appendix A) shows strong robustness against lattice dissipation but weaker robustness against atomic dissipation; nevertheless, a transfer fidelity of about 75\%.

\subsection{Entanglement Transfer For An Initial Entangled States}
\textbf{\label{section4b}}

\begin{figure}[htbp]
  \centering
    \vspace{0pt}
\includegraphics[width=1\linewidth]{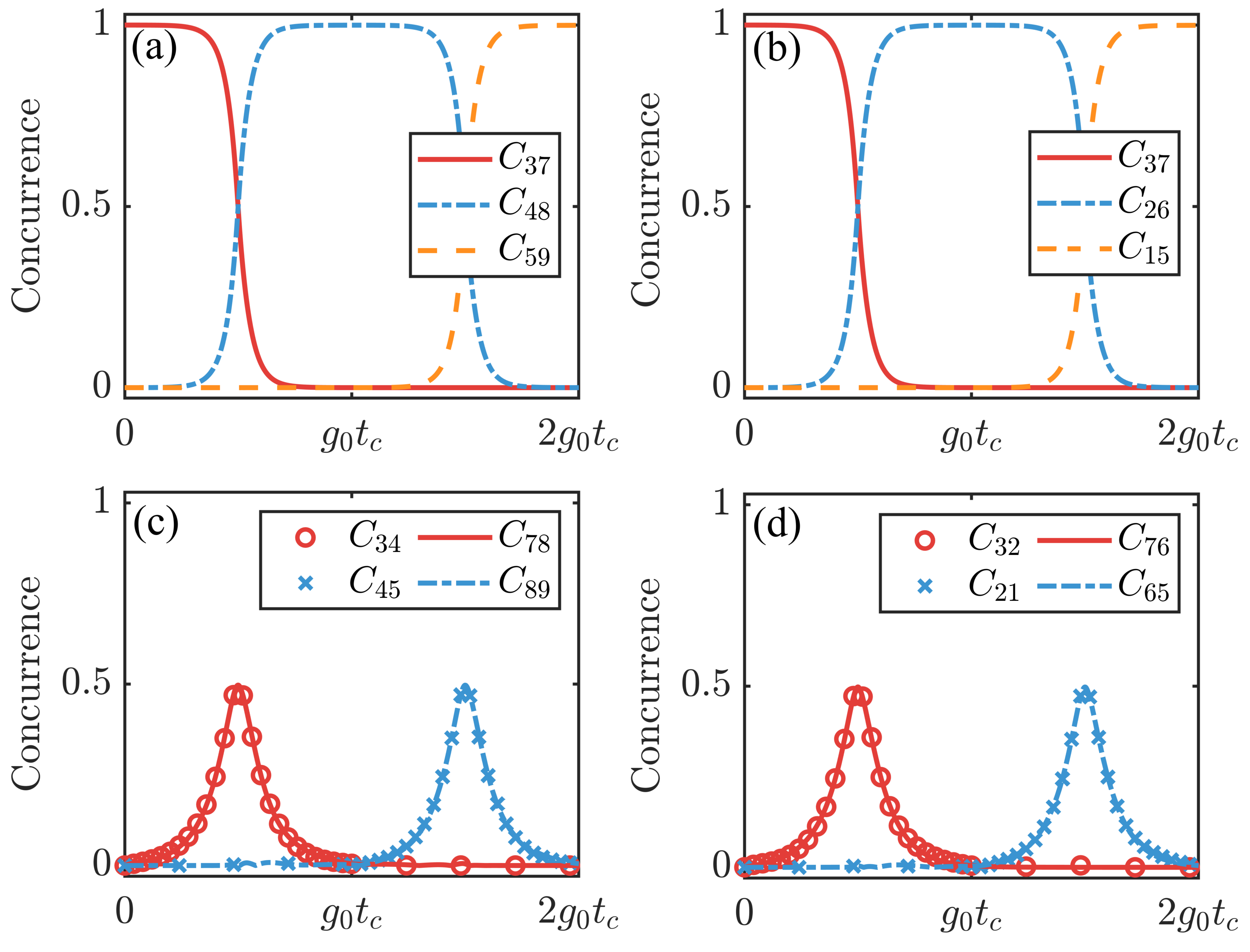}
    \caption{(a) and (b) Time evolution of the concurrence between non-neighboring atoms. (c) and (d) Time evolution of the concurrence between neighboring atoms. (a) and (c) $\varphi_l=-\pi/2$, while (b) and (d) $\varphi_l=\pi/2$. Other parameters are $N=9$, $|\Psi(0)\rangle =\left|\psi_{37}^{\pm} \right>$,  $g_0t_0=200$, $g_0\tau_{sw}=0.1$, $g_0 t_c=200.1$, $\eta/g_0=20$, $n_{2l-1}=4l-4$, $n_{2l}=4l-1$, $d_s=3$, $d_l=1$, $g_{1,3,5,7,9}(t)=g_a(t)$, and  $g_{2,4,6,8}(t)=g_b(t)$.}
    \label{FIG6}
\end{figure}

 Perfect chiral entanglement transfer can also be achieved for an entangled initial state. Considering the long-range entangled states $\left| \psi_{37} ^{\pm}\right>=\frac{1}{\sqrt{2}}(|\psi_3\rangle{\pm}|\psi_7\rangle)$ and setting the phase $\varphi_{l}=-\pi/2$ with $g_{1,3,5,7,9}(t) = g_a(t)$ and $g_{2,4,6,8}(t) = g_b(t)$. Figure~\figpanel{FIG6}{a} demonstrates sequential entanglement transfer via right-chiral spontaneous emission. During $0<g_0t<g_0t_c$, as the right-propagating photon passes, atoms 4 and 8 are simultaneously excited, yielding four-body entanglement among atoms 3,4,7 and 8. At about $t_c/2$, the concurrences $C_{34}$, $C_{78}$, $C_{37}$ and $C_{48}$ all reach 0.5 simultaneously before decreasing. Throughout this process, $C_{37}$ continues to decrease. At $t_c$, all four concurrences vanish, completing the entanglement transfer from $C_{37}$ to $C_{48}$. At this point, the four-body entanglement reduces to a two-body one and the same process subsequently transfers entanglement from $C_{48}$ to $C_{59}$. Notably, a left-propagating entanglement transfer can be achieved simply by changing the phase from $\varphi_{l}=-\pi/2$ to $\varphi_{l}=\pi/2$ (see Figs.~\hyperref[FIG6]{6(b)} and \hyperref[FIG5]{(d)}).

\begin{figure}[htbp]
  \centering
    \vspace{0pt}
\includegraphics[width=1\linewidth]{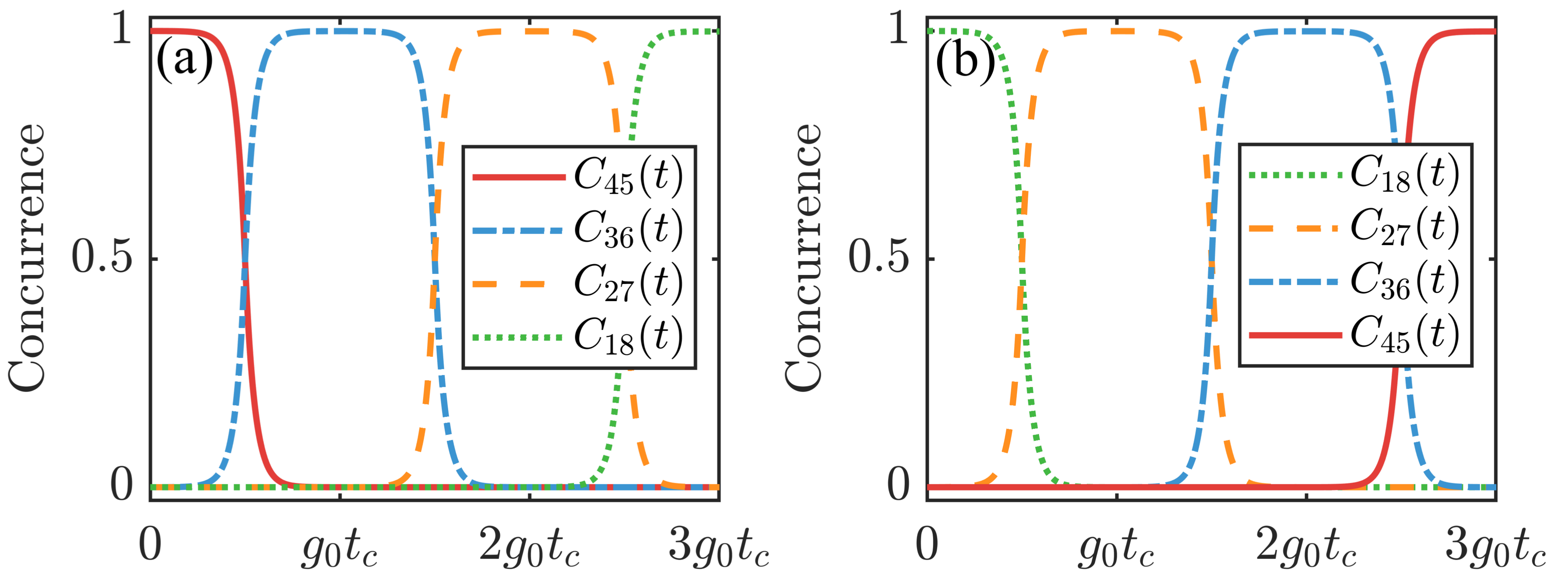}
    \caption{ (a) Time evolution of the spatial spreading of concurrence between atoms. (b) Time evolution of the spatial localization of concurrence between atoms. In (a), $g_{2,4,5,7}=g_a$, $g_{1,3,6,8}=g_b$  and $\varphi_{5,6,7,8}=-\varphi_{1,2,3,4}=-\pi/2$, with the initial state $|\Psi(0)\rangle=|\psi_{45}^{\pm}\rangle$. In (b), $g_{1,3,6,8}=g_a$, $g_{2,4,5,7}=g_b$  and $\varphi_{1,2,3,4}=-\varphi_{5,6,7,8}=-\pi/2$, with the initial state $|\psi(0)\rangle=|\Psi_{18}^{\pm}\rangle$. Other parameters are $N=8$,  $g_0t_0=200$, $g_0\tau_{sw}=0.1$, $g_0 t_c=200.1$, $\eta/g_0=20$, $n_{2l-1}=4l-4$,$n_{2l}=4l-1$, $d_s=3$, and $d_l=1$.}
    \label{FIG7}
\end{figure}

Figure~\ref{FIG7} demonstrates that the entanglement length, i.e., the distance between the two entangled atoms, can be tuned, enabling robust interconversion between long- and short-distance entanglement. In the right-chiral transfer from $C_{45}$ to $C_{18}$ (see Fig.~\figpanel{FIG7}{a}), this distance increases from one lattice spacing to seven, while in the left-chiral process (see Fig.~\figpanel{FIG7}{b}), it decreases from seven to one.

\section{back-and-forth transfer of entanglement}
\label{section5}

In the preceding discussion, the additional phase modulation $\varphi_l$ was treated as time-independent. We now consider a system of two atoms with $\varphi_l(t)$ varying periodically:
\begin{equation}
\begin{split}
\varphi _l\left( t \right) =\varphi _l\left( t+2nt_c \right) =\left\{ \begin{matrix}
	-\frac{\pi}{2},&		0\leqslant t< t_c,\\
	\frac{\pi}{2},&		t_c\leqslant t < 2t_c.\\
\end{matrix} \right. 
\end{split}
\label{EQ19}
\end{equation}
Because $\varphi_l(t)$ varies in time, the left- and right-propagating jump operators also change periodically. Setting $d_s = 3$ and using Eq.~\eqref{EQ19}, the right and left jump operators become
\begin{subequations}
\begin{align}
J_{R}^{2}(t)=
\begin{cases}
e^{ik_ad_{sw}}\sqrt{\Gamma _{R}^{2}(t)}\sigma _{2}^{-}
+\sqrt{\Gamma _{R}^{2}(t)}\sigma _{2}^{-},
& 0\leqslant t< t_c,
\\
0,
& t_c\leqslant t<2t_c,
\end{cases}
\\
J_{L}^{2}(t)=
\begin{cases}
0,
& 0\leqslant t< t_c,
\\
\sqrt{\Gamma _{L}^{2}(t)}\sigma _{2}^{-}
+e^{ik_ad_{sw}}\sqrt{\Gamma _{L}^{2}(t)}\sigma _{2}^{-},
& t_c\leqslant t< 2t_c.
\end{cases}
\end{align}
\end{subequations}
Both operators evolve with period $2t_c$: $J_{L,R}^{2}(t)=L_{J,R}^{2}\left( t+2nt_c \right)$, where $n$ is a natural number.

\begin{figure}[htbp]
  \centering
    \vspace{0pt}
    \includegraphics[width=1\linewidth]{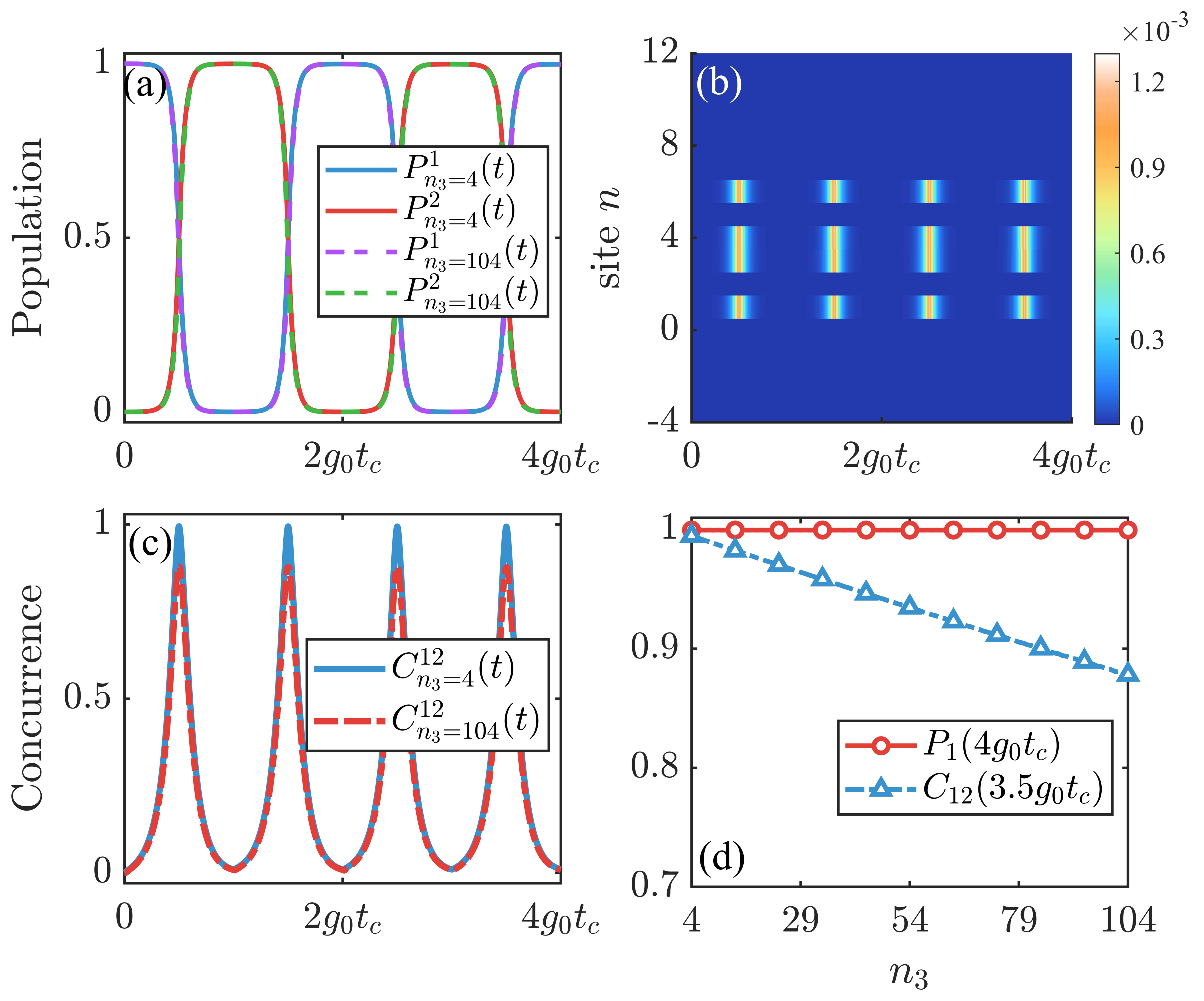}
    \caption{(a) Time evolution of the population $P_l(t)$ for different values of $n_3$. (b) Time evolution of the field intensity distribution $|c_j(t)|^2$ for $n_3=4$. (c) Time evolution of the concurrence between atoms for different values of $n_3$. (d) Population of the state $|\psi_1\rangle$ at the time $4g_0t_c$ and the concurrence $C_{12}$ at the time $3.5g_0t_c$ as functions of $n_3$. Here $\varphi_1(t) = \varphi_2(t)$, and the dynamics follow Eq.~\eqref{EQ19}.  For $n_3=4$, $g_0\tau_{sw}=0.1$; for $n_3=104$, $g_0\tau_{sw}=2.6$. Other parameters are $N=2$,  $|\Psi (0)\rangle =|\psi_1\rangle$, $g_0t_0=200$, $g_0 t_c=g_0(t_0+\tau_{sw})$, $ \eta/g_0=20$, $n_{1}=0$, $n_{2}=3$, $n_4=n_3+3$, $d_s=3$, $d_w=n_3-n_{2}$, $g_1(t) = g_a(t)$, and $g_2(t)= g_b(t)$.}
    \label{FIG8}
\end{figure}

During $0<t<t_c$, atoms emit only into the right-propagating mode, and the dark-state condition applies to $J_{R}^{2}(t)|\Psi(t)\rangle$. During $t_c<t<2t_c$, emission is only into the left-propagating mode, with $J_{L}^{2}|\Psi(t)\rangle$ satisfying the dark-state condition. This yields 
\begin{equation}
\begin{cases}
e^{ikd_{sw}}\sqrt{\Gamma _1(t)}u_1(t)
+\sqrt{\Gamma _2(t)}u_2(t) =0,
& 0\leqslant t<t_c,
\\
\sqrt{\Gamma _1(t)}u_1(t)
+e^{ikd_{sw}}\sqrt{\Gamma _2(t)}u_2(t) =0,
& t_c\leqslant t< 2t_c,
\end{cases}
\label{EQ21}
\end{equation}
which also oscillates with period of $2t_c$. Under this constraint and with $g_1(t)=g_a(t)$ and $g_2(t)=g_b(t)$, the system undergoes coherent state transfer from $|\psi_1\rangle$ to $|\psi_2\rangle$ for $0<g_0t<g_0t_c$. and back from $|\psi_2\rangle$ to $|\psi_1\rangle$ for $g_0t_c<g_0t<2g_0t_c$. These two transitions alternate periodically (see Fig.~\figpanel{FIG8}{a}), as also seen in the lattice excitation probabilities (see Fig.~\figpanel{FIG8}{b}). Because entanglement dynamics follows state transfer, each complete transition corresponds to one cycle of entanglement generation and decay (see Fig.~\figpanel{FIG8}{c}).

In Figs.~\hyperref[FIG8]{\ref*{FIG8}(a)}-\hyperref[FIG8]{(c)}, the interatomic distance $d_{sw}$ is small, so the propagation delay $\tau_{sw}$ is short. Increasing the distance (see Fig.~\figpanel{FIG8}{d}) still yields high-quality long-range entanglement transfer with nearly perfect state transfer, because the delay effect is already absorbed into the coupling coefficient $g_l(t)$. The periodic state transfer closely resembles the decoherence-free interactions in conventional braided giant-atom configurations.

Compared with conventional braided giant-atom architectures, the present scheme enables nearly perfect sustained quantum state exchange even in the presence of strong non-Markovian effects. Remarkably, such long-distance sustained exchange can be achieved using only two spatially separated giant atoms, without resorting to complex braided coupling networks.

\section{Conclusions}
\label{section6}
We have shown that controllable chiral propagation of entanglement can be achieved in a system of giant atoms coupled to a one‑dimensional waveguide, enabling perfect unidirectional sequential transfer of quantum states and their corresponding entanglement. By tuning an additional phase, selective transfer between non‑adjacent atoms becomes accessible. Moreover, the entanglement length (the distance between two entangled atoms) can be varied dynamically, allowing robust interconversion between long‑ and short‑range entanglement during transfer. Engineering a periodic piecewise phase modulation further enables two distant giant atoms to exhibit persistent, nearly lossless state exchange (alternating back and forth) and to sustain stable entanglement even under non‑Markovian dynamics. This behaviour mimics that of conventional braided giant‑atom configurations but avoids their structural constraints and propagation‑delay limitations. Furthermore, the system spontaneously evolves into and remains in a dark state, ensuring high‑fidelity directional transfer of quantum states and entanglement. Our scheme thus provides a scalable route towards continuous, long‑distance entanglement transport and robust state exchange in quantum networks.

\section*{Acknowledgments}

We thank Dr. Lei Du for fruitful discussions. This work is supported by the National Natural Science Foundation of China  (Grants No. 12564047, No. 11874004, No. 11204019, and No. 12564048), Hainan Province Flexible Talent Introduction Collaborative Innovation Center (Yu Changbin), and the specific research fund of The Innovation Platform for Academicians of Hainan Province (No: YSPTZX202407).

\appendix
\renewcommand{\appendixname}{\MakeUppercase{APPENDIX}}

\section{\MakeUppercase{Effect of Dissipation on Entanglement Transfer}}
\label{appendix A}

\begin{figure*}[htbp]
  \centering
    \vspace{0pt}
    \includegraphics[width=0.98\textwidth]{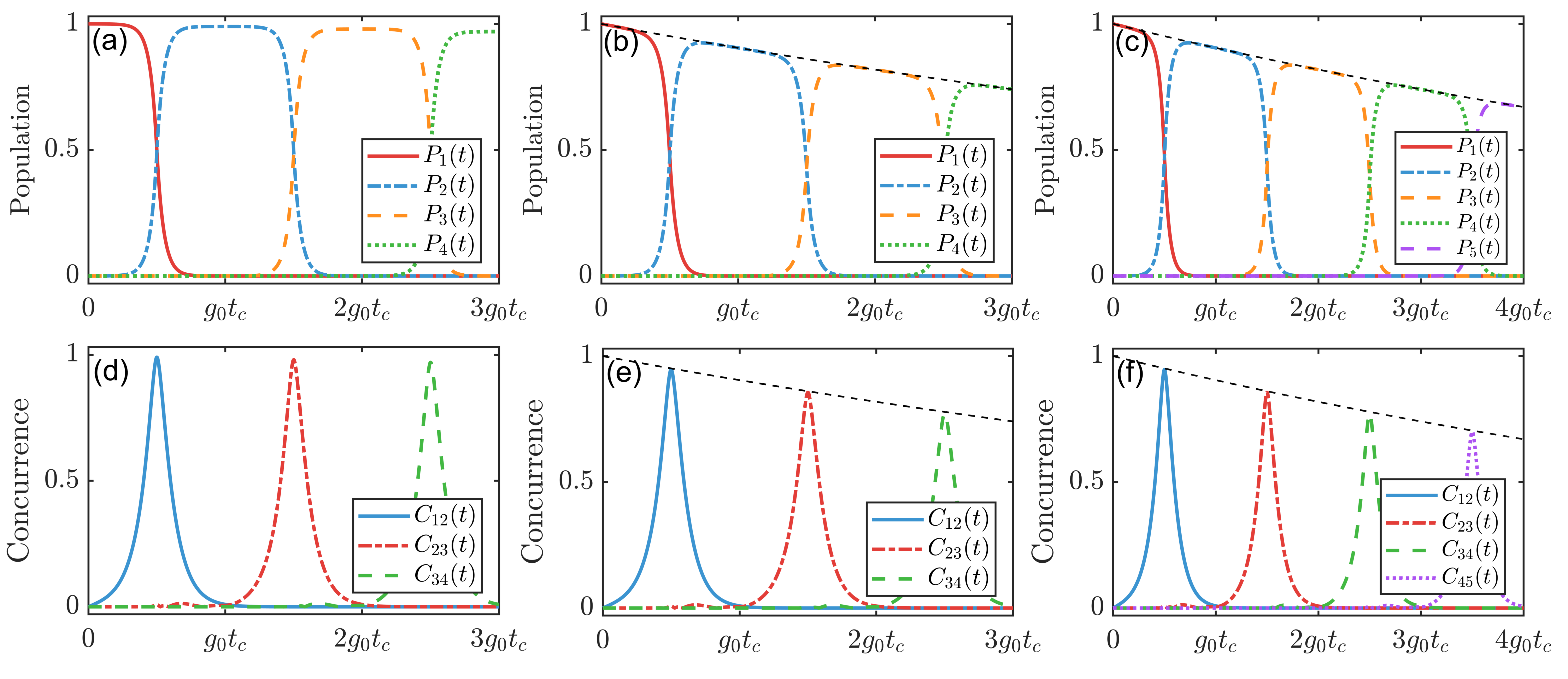}
    \caption{(a)–(c) Time evolution of the population $P_l(t)$ associated with state $|\psi_l\rangle$. (d)–(f) Time evolution of the concurrence between neighboring atoms. The gray dashed line indicates the dissipation curve jointly followed by the population probability and the concurrence measure, whose decay behavior satisfies $y=e^{-2\kappa_at}$. Case I: $\kappa_0/g_0 = 0.05$, $\kappa_a = 0$, $g_0t_0= 200$. Case II: $\kappa_0 = 0$, $\kappa_a/g_0 = 2.5\times 10^{-4}$, $g_0t_0= 200$. Case III: $\kappa_0 = 0$, $\kappa_a/g_0 = 2.5\times 10^{-4}$, $g_0t_0 = 150$. Other parameters are $|\Psi (0)\rangle =|\psi_1\rangle$, $g_0\tau_{sw}=0.1$, $g_0 t_c=g_0(t_0+\tau_{sw})$, $\eta/g_0=20$, $g_{1,3,5}(t)=g_a(t)$, $g_{2,4}(t)=g_b(t)$, $n_{2l-1}=4l-4$, $n_{2l}=4l-1$, $d_s=3$, $d_l=1$, and $\varphi_l=-\pi/2$.}
    \label{SFIG1}
\end{figure*}

Fig.~\ref{FIG5} shows an ideal atom–waveguide system without external dissipation, where the waveguide is the sole reservoir. With appropriate parameters, near‑perfect state and entanglement transfer is achievable. In realistic systems, however, atomic and lattice dissipation are unavoidable. The Hamiltonians then become
\begin{equation}
\begin{aligned}
H_a &= \sum_{l=1}^N (\omega_a - i\kappa_a)\, \sigma_{l}^{+} \sigma_{l}^{-}, \\
H_w &= \sum_j (\omega_0 - i\kappa_0)\, a_{j}^{\dagger} a_j 
- \sum_j \eta \left( a_j a_{j+1}^{\dagger} + \mathrm{H.c.} \right),
\end{aligned}
\end{equation}
where $\kappa_a$ and $\kappa_0$ are atomic and lattices decay rates, respectively.

Case I ($\kappa_0/g_0 = 0.05$, $\kappa_a = 0$, $g_0t_c= 200$). Figure~\figpanel{SFIG1}{a} and Figure~\figpanel{SFIG1}{d} show that both state and entanglement transfer remain at high levels, indicating strong robustness against lattice dissipation.

Case II ($\kappa_0 = 0$, $\kappa_a/g_0 = 2.5\times 10^{-4}$, $g_0t_c = 200$). Figure~\figpanel{SFIG1}{b} and Figures~\figpanel{SFIG1}{e} reveal a marked drop in fidelities. The decay of atomic population and concurrence follows $y=e^{-2\kappa_at}$ (grey dashed line in Fig.~\ref{SFIG1}). Within $0\le g_0t \le 3g_0t_c$, however, fidelities stay relatively high. Comparing Case I with Case II shows that atomic dissipation is the main limiting factor.

Case III ($\kappa_0 = 0$, $\kappa_a/g_0 = 2.5\times 10^{-4}$, $g_0t_c = 150$). Shortening the single-transfer window allows more transfer cycles within the same total time $0\le g_0t \le 4g_0t_c$ (see Fig.~\figpanel{SFIG1}{c} and Fig.~\figpanel{SFIG1}{f}). By comparing the time parameters in Cases II and III, it is evident that although the number of quantum-state and entanglement transfer events differs between the two scenarios, the total evolution time is nearly identical, and the corresponding fidelities are essentially the same, both following the same exponential decay law.

This behaviour arises because only one effective transfer process occurs at any given time; dissipation involves at most two atoms simultaneously. Consequently, the decay of populations and concurrence, as well as the final transfer fidelity, depend only on the total evolution time, not on the number of atoms or transfer events.

\twocolumngrid
\bibliography{GA_ref.bib}
\end{document}